\documentclass[amsmath,amssymb]{revtex4-2}

\usepackage[breaklinks=true,colorlinks,citecolor=blue,linkcolor=blue,urlcolor=blue]{hyperref}
\usepackage{float}
\usepackage{mathrsfs,xcolor,latexsym,graphicx,array,amsmath,amsfonts,amssymb,tabularx,placeins}

\makeatletter
\newcommand*{\rom}[1]{\expandafter\@slowromancap\romannumeral #1@}
\makeatother

\begin{document}
\title{Theory of Spiral Magnetism in Weyl semimetal SmAlSi}

\author{Lasin Thaivalappil}
\affiliation{Department of Physics, National Sun Yat-sen University, Kaohsiung 804, Taiwan}

\author{Rahul Verma}
\affiliation{Department of Condensed Matter Physics and Materials Science, Tata Institute of Fundamental Research, Colaba, Mumbai 400005, India}

\author{Hsin Lin}
\affiliation{Institute of Physics, Academia Sinica, Taipei 11529, Taiwan}

\author{Bahadur Singh}
\email{Contact author: bahadur.singh@tifr.res.in}
\affiliation{Department of Condensed Matter Physics and Materials Science, Tata Institute of Fundamental Research, Colaba, Mumbai 400005, India}

\author{Shin-Ming Huang}
\email{Contact author: aesthin@gmail.com}
\affiliation{Department of Physics, National Sun Yat-sen University, Kaohsiung 804, Taiwan}
\affiliation{Center for Theoretical and Computational Physics, National Sun Yat-sen University, Kaohsiung 80424, Taiwan}
\affiliation{Physics Division, National Center for Theoretical Sciences, Taipei, 10617, Taiwan}

\begin{abstract}
Recent neutron scattering and thermodynamic measurements suggest that Weyl electrons in the emergent Weyl semimetal SmAlSi mediate unconventional magnetic interactions and induce spiral magnetic order. In this work, we investigate the nature of these interactions by modelling long-range $f-f$ exchange mediated by itinerant $d$ electrons via the Ruderman-Kittel-Kasuya-Yosida (RKKY) mechanism, employing a material-specific tight-binding Hamiltonian obtained from first-principles calculations. The magnetic susceptibility is derived from the spin-spin correlation function based on the random phase approximation. Our results demonstrate that Fermi-surface nesting alone cannot account for the experimentally observed magnetic modulation at the wave vector (1/3, 1/3, 0); however, incorporating appropriate antiferromagnetic exchange interactions among the $d$ electrons yields the correct propagation vector. The spin-texture analysis reveals a configuration that is close to a cycloidal spin structure, preserving combined glide and time-reversal symmetry, and reflecting an intricate competition between inter- and intra-sublattice interactions in SmAlSi.
\end{abstract}

\maketitle

\textbf{Introduction.~}
Magnetic Weyl semimetals (MWSMs) have attracted considerable interest in condensed matter physics due to their exotic quantum properties and potential for next-generation device applications~\cite{bib1,bib2,bib13,bib16,bib29,bib33,bib41,bib11,bib20}. In non-centrosymmetric MWSMs, broken inversion symmetry gives rise to Weyl nodes with well-defined chirality, while magnetic order fine-tunes their separation in momentum space~\cite{bib6,bib15,bib37,bib42}. Recent studies on the RAlSi family (R = Ce, Pr, Nd, and Sm) reveal that, in addition to hosting chiral Weyl electrons near the Fermi level, these compounds exhibit distinct magnetic ground states ~\cite{bib46}. CeAlSi is a noncollinear ferromagnet with strong magnetocrystalline anisotropy, which progressively weakens toward SmAlSi. These materials not only exhibit unconventional Hall and transport responses driven by Weyl electrons but also provide a compelling platform for studying the interplay between magnetic interactions and topological electronic states across different magnetic regimes. 

Among RAlSi, SmAlSi uniquely hosts multiple Weyl nodes and associated remnant electron-hole pockets at the Fermi level. Neutron scattering, transport, and thermodynamic measurements reveal a spiral magnetic ground state with an ordering vector $\mathbf{Q} = (\frac{1}{3}, \frac{1}{3}, 0)$~\cite{bib35,bib36}, which cannot be explained through the nesting of Fermi surface pockets alone. These findings point towards an active role of Weyl electrons in mediating spiral magnetic order through indirect Ruderman-Kittel-Kasuya-Yosida (RKKY) interactions. However, a detailed theoretical understanding of the specific magnetic interactions remains lacking and requires detailed investigation using material-specific modeling.

In this work, we investigate how the Weyl electron dispersion influences the RKKY interaction in SmAlSi by modeling the magnetic behavior and spin ordering based on a first-principles-derived electronic structure. We examine the magnitude and momentum dependence of the RKKY interaction in the presence of Weyl nodes and Fermi surface nesting, taking into account the role of interactions among itinerant electrons. The RKKY interaction is computed within the random phase approximation (RPA)~\cite{bib38, bib39}, considering three short-range superexchange parameters among the itinerant $d$ electrons: onsite ($J_1$), inter-sublattice ($J_2$), and intra-sublattice ($J_3$). A magnetic phase diagram is constructed by optimizing a mean-field single-$q$ ansatz, which reveals that the experimentally observed spiral order with wavevector $\left(\frac{1}{3}, \frac{1}{3}, 0\right)$ is favored when both $J_2$ and $J_3$ are antiferromagnetic, and $J_1$ is ferromagnetic, consistent with expectations for strongly correlated materials.

The optimal magnetic state is a single-$q$ spiral exhibiting combined $\mathcal{T}\mathcal{G}$ symmetry, where the glide plane $\mathcal{G}$ is perpendicular to $\mathbf{Q}$. The magnetic texture resembles a hybrid helical–cycloidal structure, but the presence of $\mathcal{T}\mathcal{G}$ symmetry brings it closer to a cycloidal configuration. This magnetic state, dictated by the material’s magnetic space group that couples spatial and time-reversal symmetries, underscores the crucial role of crystal symmetry in stabilizing unconventional magnetic orders.

We also employed the Luttinger–Tisza method, which approximates the ground-state energy by selecting the lowest eigenvalue of the interaction matrix at each momentum point. However, this method becomes unreliable when off-diagonal intra- and inter-sublattice interactions are significant. In this material, although the Luttinger–Tisza approach still yields the energy minimum at $\mathbf{Q}$, it fails to capture the correct magnetic structure.

\begin{figure}[ht!]
\centering
\includegraphics[width=10 cm]{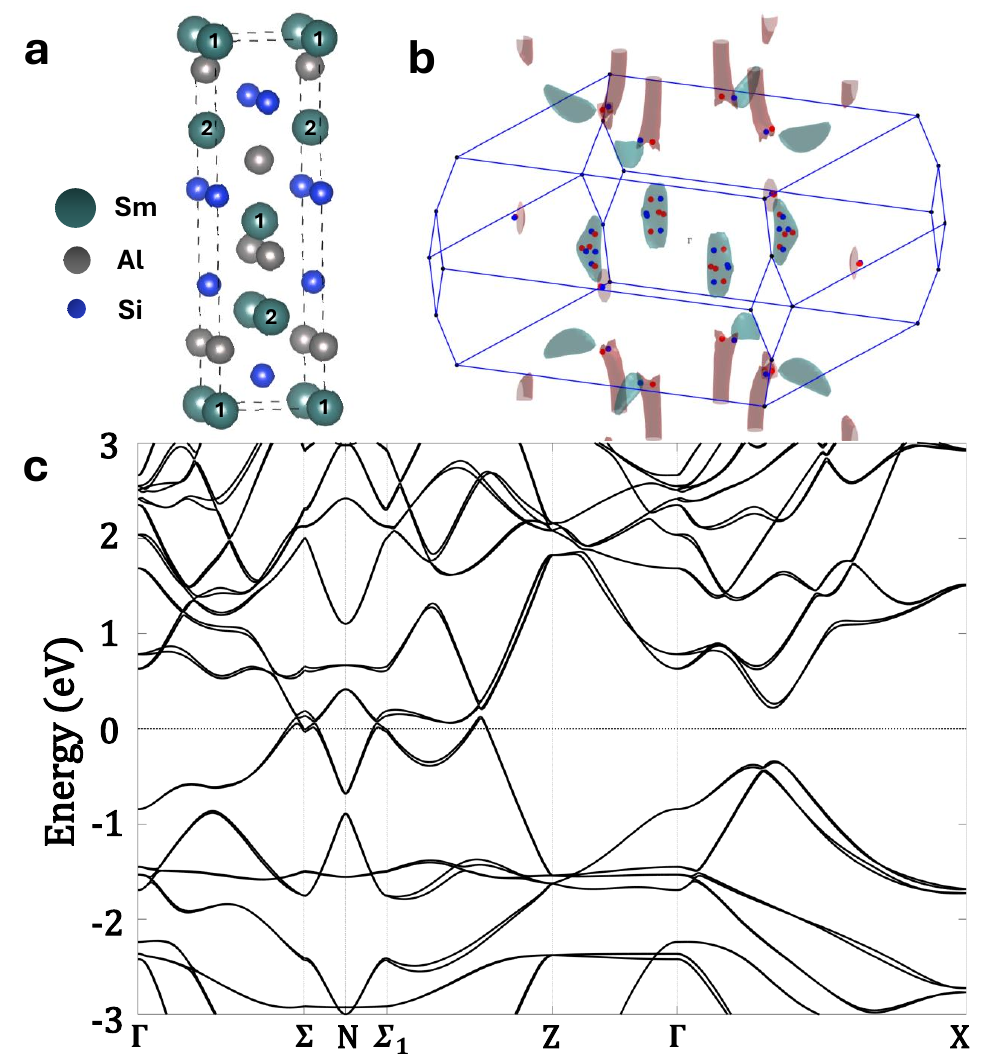}
\caption{(a) Crystal structure of SmAlSi, showing the two Sm sublattices. (b) Fermi surface with electron and hole pockets shown in red and blue, respectively. The distribution of Weyl nodes in the Brillouin zone is indicated, with red for positive ($C_w = +1$) and blue for negative ($C_w = -1$) chiralities. (c) Calculated nonmagnetic band structure of SmAlSi including spin-orbit coupling. The dotted line marks the Fermi energy. }
\label{fig:1}
\end{figure}

\textbf{Crystal and electronic structure.~}
SmAlSi crystallizes in a body-centered tetragonal structure (see Fig. \ref{fig:1}) and belongs to the acentric space group $I4_1md$ (no. 109). It lacks inversion symmetry and contains two mirror planes, $m_{100}$ and $m_{010}$, along with two glide planes, $\mathcal{G}_1 = \{ m_{1 1 0} | 0 \frac{1}{2} \frac{1}{4} \}$ and $\mathcal{G}_2=\{m_{1 \bar{1} 0} | \frac{1}{2} 0 \frac{3}{4}\}$. It contains $\mathcal{S}_1  = \{4_{001}|0\frac{1}{2} \frac{1}{4} \}$ screw rotation, which can be generated through a combination of a mirror and a glide operation. The crystal structure contains two magnetic sublattices related by either a screw or a glide symmetry. We define sublattice-1 and sublattice-2 based on the positions of Sm atoms located at $(0, 0, 0)$ and $(\frac{1}{2}, 0, \frac{1}{4})$, respectively (see Fig. \ref{fig:1}(a)).

\begin{figure*}[t!]
\centering
\includegraphics[width=1\linewidth]{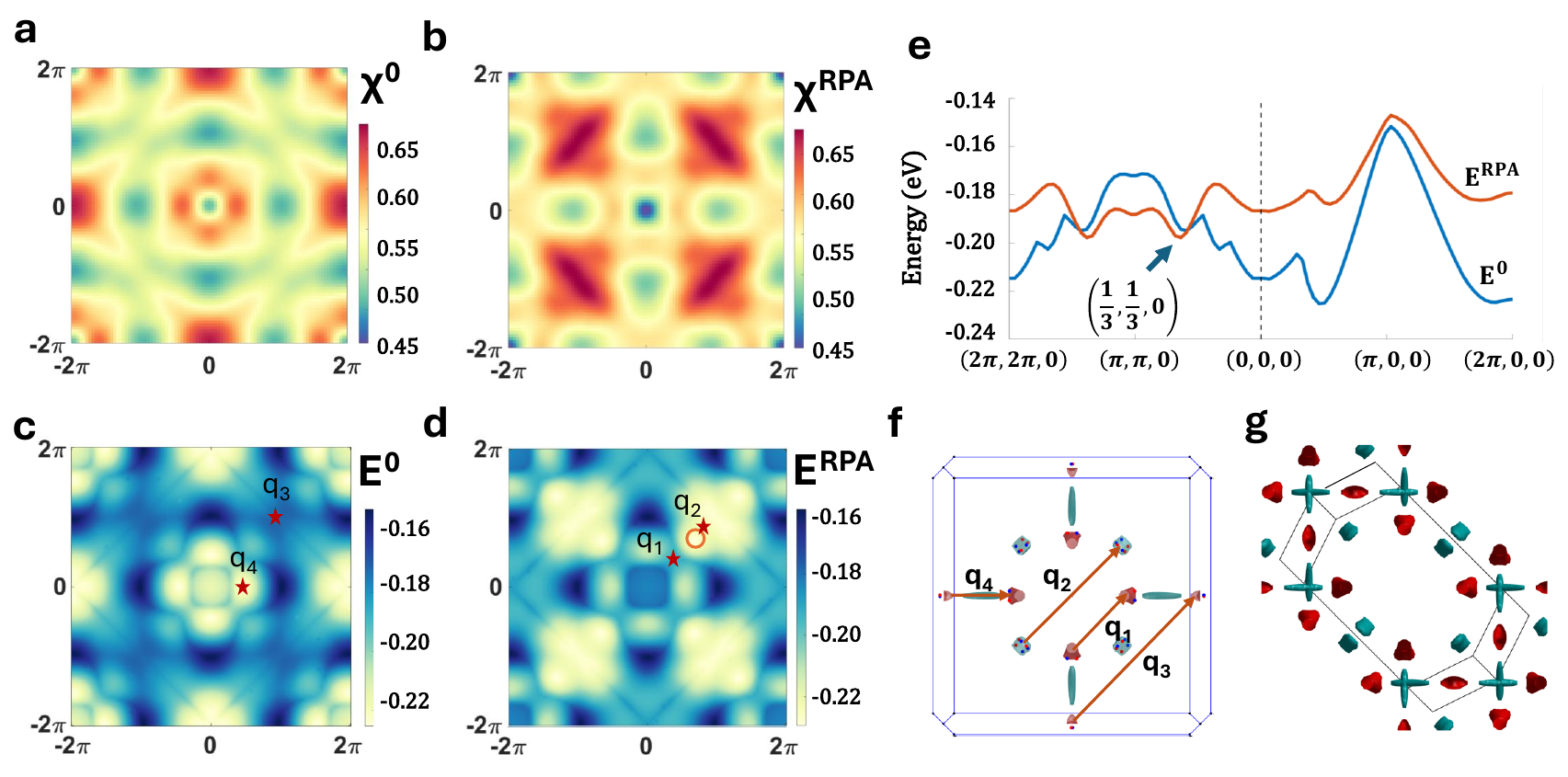}
\caption{(a), (b) Half the trace of the susceptibility ($\frac{1}{2}\sum_{l,\alpha} \chi^{ll}_{\alpha \alpha } $) in the $ q_z=0$ plane for $\tilde{\chi}^{\text{0}}$ and for $\tilde{\chi}^{\text{RPA}}$, respectively. (c), (d) Optimal variational energy density for the single-$q$ state in the $ q_z=0$ plane based on $\tilde{\chi}^{\text{0}}$ ($E^{0}$) and on $\tilde{\chi}^{\text{RPA}}$ ($E^{\text{RPA}}$), respectively. The global energy minimum is denoted by the circle in (d). (e) Variational energy $E^0$ (blue) and $E^{\text{RPA}}$ (red) along the (110) and (100) directions. The minimum of $E^{\text{RPA}}$ indicated by the arrow appears at $(\frac{1}{3}, \frac{1}{3}, 0)$. (f) Top view of the Fermi surface with the Weyl nodes. The arrow represents promising nesting vectors $q_1$, $q_2$, $q_3$, and $q_4$. The locations of $q_1 = (0.22, 0.22, 0)$, $q_2 = (0.36, 0.36, 0)$, $q_3 = (0.49, 0.49, 0)$, and $q_4 = (0.25, 0, 0)$ are marked by stars in (c) and (d). (g) Folded Fermi surfaces by $\mathbf{Q}$ with the reduced Brillouin zone edges. In (b), (c), and (d), values of $J$'s are seen in the main text.}
\label{fig:2}
\end{figure*}

The electronic structure of SmAlSi is calculated using the density functional theory framework as implemented in the Vienna ab initio simulation package (VASP)~\cite{Hohen,Bloch,Kresse1996,Kresse1999}. To model the nonmagnetic state, the Sm $4f$ electrons are treated as core states, and a Wannier tight-binding Hamiltonian, $H_t$, is generated. We use the experimental lattice parameters ($a = b = 4.153$ Å, $c = 14.408$ Å)~\cite{bib35}, and relax the internal atomic positions until the residual forces on each atom are less than $10^{-2}$ eV/\r{A}, employing a high-energy cutoff and $k$-point mesh (see Supplemental Materials (SMs) for details). The calculated band structure of the nonmagnetic phase of SmAlSi is shown in Fig.~\ref{fig:1}(c), with the Fermi energy set to zero. The bands near the Fermi level resemble those of other compounds in the RAlSi family and exhibit Weyl nodal crossings as illustrated in Fig.~\ref{fig:1}(b). Due to crystal symmetry, each Weyl node has a mirror-reflected counterpart with opposite chirality, while screw rotations generate additional Weyl nodes with the same chirality. In total, there are 28 pairs of Weyl nodes in the first Brillouin zone. Depending on whether a Weyl node lies below or above the Fermi energy, it is enclosed by either an electron or a hole pocket, as seen in Fig.~\ref{fig:1}(b). A top-down view of the Fermi surfaces is provided in Fig.~\ref{fig:2}(f).

An analysis of the Fermi surface reveals that specific segments are interconnected by characteristic nesting vectors $q_1$, $q_2$, $q_3$, and $q_4$, as shown in Fig.\ref{fig:2}(f), with their corresponding positions marked by stars in Fig.~\ref{fig:2}(c) and (d). Although these vectors can lower the energy of the system by introducing electronic or magnetic instabilities, our analysis shows that none of them match the $\mathbf{Q}$ vector marked by the red circle in Fig.~\ref{fig:2}(d) (see discussion below).  This scenario is clearer in Fig.~\ref{fig:2}(g), which shows the folding of the Fermi surface along the vector $\mathbf{Q}$. A very small overlap of Fermi pockets is observed, indicating that Fermi surface nesting has minimal impact on the ground state.

\textbf{Electron correlation and susceptibility.~}
Since the Fermi pockets in SmAlSi do not exhibit nesting at the experimentally observed $\mathbf{Q}$ vector, nesting alone cannot account for the observed antiferromagnetic order. These pockets are also relatively small, leading to a low density of states at the Fermi level and, consequently, weak conventional RKKY interactions. However, because the pockets surround the Weyl nodes, the RKKY interactions mediated by Weyl fermions remain significant. Unlike conventional RKKY interactions, which are primarily governed by the Fermi-level DOS, the Weyl-mediated RKKY interaction receives substantial contributions from high-energy states, leading to an ultraviolet cutoff dependence: $J \propto \Lambda^2$~\cite{bib40}. To include the RKKY interaction, we incorporate both $f$-$d$ and $d$-$d$ interactions into the first-principles Wannier tight-binding Hamiltonian $H_t$. The full Hamiltonian is

\begin{equation} \label{eq:1}
    H_0 = H_t + H_{fd} + H_{dd},
\end{equation}
where $H_{fd}$ and $H_{dd}$ describe the $f$-$d$ and $d$-$d$ interactions, respectively, and given as
\begin{equation} \label{eq:2-1}
\small
    H_{fd} = -J_H \sum_{\mathbf{r}} \sum_{l=1,2} \vec{S}^{(l)}(\mathbf{r}) \cdot \vec{s}^{(l)}(\mathbf{r}), 
\end{equation}
and 
\begin{align} \label{eq:2-2}
   \small
    \begin{split}
         H_{dd} =& J_1 \sum_{\mathbf{r}} \sum_{l}\vec{s}^{(l)}(\mathbf{r}) \cdot \vec{s}^{(l)}(\mathbf{r}) + J_2  \sum_{\langle \mathbf{r},\mathbf{r}' \rangle} \sum_{l} \vec{s}^{(l)}(\mathbf{r}) \cdot \vec{s}^{(\bar{l})}(\mathbf{r}') \\
         &+ J_3 \sum_{\langle \langle  \mathbf{r},\mathbf{r}' \rangle \rangle} \sum_{l} \vec{s}^{(l)}(\mathbf{r}) \cdot \vec{s}^{(l)}(\mathbf{r}') .
    \end{split}
\end{align}

Here, $\vec{S}^{(l)}$ and $\vec{s}^{(l)}$ denote the total spins of $f$ and $d$ orbitals, respectively, on an Sm atom at sublattice $l$. The vector $\mathbf{r}$ runs over the points of the Bravais lattice. $\langle \mathbf{r},\mathbf{r}' \rangle$ ($\langle \langle \mathbf{r},\mathbf{r}' \rangle \rangle$) denotes nearest (next-nearest) neighbour distance with Sm atoms belonging to different (same) sublattices. For example, a Sm atom in sublattice 1 is tetrahedrally coordinated by four nearest neighbors from sublattice 2, while the nearest intra-sublattice neighbors lie in the same $x$–$y$ plane at distance $\pm |\vec{x}|$ or $\pm |\vec{y}|$. In $H_{dd}$, three interactions are considered: the on-site $J_1$, the nearest inter-sublattice $J_2$, and the nearest intra-sublattice $J_3$ interactions.
 
By applying second-order perturbation theory, or equivalently integrating out the degrees of freedom of the $d$ electrons, the effective Hamiltonian describing RKKY interaction is generated as
\begin{equation} \label{eq:4}
    H_{\text{eff}}    = \sum_{\mathbf{q}} \sum_{ \substack{\alpha, \beta \\ \in \{x,y,z\} } } \sum_{l,l' } \mathcal{J}_{ \alpha \beta}^{l l'}(\mathbf{q}) S_{ \alpha}^{(l)}(-\mathbf{q}) S_{ \beta}^{(l')}(\mathbf{q}),
\end{equation}
where $\vec{S}^{(l)}(\mathbf{q}) = \frac{1}{\sqrt{N}}\sum_{\mathbf{r}} e^{-i \mathbf{q} \cdot \mathbf{r}}\vec{S}^{(l)}(\mathbf{r})$ with $N$ denoting the number of unit cells. The spin components (subscripts) and sublattice indices (superscripts) are retained for clarity. The effective RKKY interaction arises through three processes: (i) an $f$ orbital interacts with itinerant $d$ orbitals near the Fermi surface via $H_{fd}$, (ii) the itinerant $d$ orbitals interact among themselves through $H_{dd}$, and (iii)  the $d$ orbitals couple to an $f$ orbital on another site. The resulting effective exchange $\mathcal{J}$ is given as
\begin{equation}
    \mathcal{J}_{ \alpha \beta}^{l l'}(\mathbf{q}) = -J_H^2 \chi_{ \alpha \beta}^{l l'}(\mathbf{q}), \label{eff_J}
\end{equation}
where $\chi_{ \alpha \beta}^{l l'}(\mathbf{q})$ is the magnetic susceptibility of the itinerant electrons, and the Hund's coupling $J_H$ is set to 1 eV.
\begin{align} \label{eq:5}
\chi_{\alpha \beta}^{l l'}(\mathbf{q}) = \langle \langle   s^{(l)}_{\alpha}(\mathbf{q},\tau) s^{(l')}_{ \beta }(-\mathbf{q},0) \rangle \rangle_{\omega = 0},
\end{align}
Here, spin operator $s_{\alpha}$ for itinerant $d$ electrons is 
$s_{ \alpha}^{(l)}(\mathbf{q}) = \frac{1}{2\sqrt{N}} \sum_{\mathbf{k},s} \sum_{\sigma \sigma' = \uparrow , \downarrow} c^{(l) \dagger}_{\mathbf{k},s,\sigma} \sigma^\alpha_{\sigma  \sigma'} c_{\mathbf{k}+\mathbf{q},s, \sigma'}^{(l)}$, where $s$ labels the $d$ orbitals. 

To evaluate the magnetic susceptibility, we use RPA, where the susceptibility is expressed as
\begin{equation} \label{eq:6}
   \tilde{\chi}^{\text{RPA}}(\mathbf{q}) = \left[ \tilde{\mathbb{I}} + \tilde{\chi}^{\text{0}}(\mathbf{q}) \tilde{V}(\mathbf{q}) \right]^{-1} \tilde{\chi}^{\text{0}}(\mathbf{q}),   
\end{equation}
with $\Tilde{\chi}^{\text{0}}$ is the bare susceptibility. Here, $\tilde{\chi}^{\text{RPA}}(\mathbf{q})$ is a $6 \times 6$ matrix that includes two sublattices and three spin components, and $\Tilde{\mathbb{I}}$ is the identity matrix. Arranging the matrices of the susceptibility and the interaction vortex as

\begin{equation}    \label{eq:7}
    \Tilde{\chi}^{\text{RPA}} = \begin{bmatrix} \hat{\chi}_{11} & \hat{\chi}_{12} \\ \hat{\chi}_{21} & \hat{\chi}_{22}
    \end{bmatrix},
\end{equation}
and
\begin{equation}    \label{eq:8}
\Tilde{V} = \begin{bmatrix}
\hat{V}_{11} & \hat{V}_{12} \\
\hat{V}_{12}^{\dagger} & \hat{V}_{22}
\end{bmatrix}.
\end{equation}
where $11$ and $22$ ($12$ and $21$) represent intra-sublattice (inter-sublattice) terms. Each term is a $3\times 3$ matrix acting on three spin components. 
Since the spin interactions among $d$ electrons are of isotropic Heisenberg-type, the $\hat{V}$ matrices are diagonal as $\hat{V}_{l l'}=V_{l l'}\hat{\mathbb{I}}$ with
\begin{align}   \label{eq:9}
    V_{\text{11}} = V_{22} =J_1 + 2 J_3 \left[ \cos(\mathbf{q} \cdot \hat{x}) +\cos(\mathbf{q} \cdot \hat{y})\right],
\end{align}
and
\begin{align}   \label{eq:10}
\small
     V_{\text{12}} &= 2 J_2 \left[ e^{i \frac{\mathbf{q} \cdot \hat{z}}{2}}\cos\left(\frac{\mathbf{q} \cdot \hat{x}}{2}\right) + e^{- i \frac{\mathbf{q} \cdot \hat{z}}{2}} \cos\left(\frac{\mathbf{q} \cdot \hat{y}}{2}\right) \right]
\end{align}
 The detailed calculation of the susceptibility and the complete inter- and intra-sublattice susceptibility is provided in the Supplemental Materials.

Using the effective exchange interaction in Eq.~\eqref{eff_J}, the magnetic ground state of the $f$ spin system can be determined. We employ both the mean-field ansatz method and the Luttinger-Tisza method to determine the optimal modulation vector. The mean-field ansatz method works with the strict spin constraint, i.e., fixing the magnitude of each spin to a constant, $\left| \vec{S}^{(l)}(\pmb{r}) \right| = S$. 
Because it does not approximate the spin constraint, this method yields more reliable results. 
The mean-field ansatz we consider includes the single-$q$ and double-$q$ ones as follows respectively~\cite{bib48},
 \begin{align}   \label{eq:11}  
\vec{S}^{(l)}_{1Q}(\pmb{r}) &= S \left(\hat{e}_1^{(l)} \cos(\mathbf{q} \cdot \pmb{r}) + \hat{e}_2^{(l)} \sin(\mathbf{q} \cdot \pmb{r}) \right),
\\
\vec{S}_{2Q}^{(l)}(\pmb{r}) &= S \Bigg( \sqrt{1 - b^{2} \sin^2(\mathbf{q}_2 \cdot \pmb{r})} \left[ \hat{e}_1^{(l)} \cos(\mathbf{q}_1 \cdot \pmb{r}) \right. \nonumber \\
&\qquad \left. - \hat{e}_2^{(l)} \sin(\mathbf{q}_1 \cdot \pmb{r}) \right] + \hat{e}_3^{(l)} b \sin(\mathbf{q}_2 \cdot \pmb{r}) \Bigg),  \nonumber 
\end{align}
    where $\hat{e}_i^{(l)}~(l=1,2)$ and $b$ are variational parameters; the former are orthonormal unit vectors defining the spin orientation on sublattice $l$, and the latter describes the magnitude of the second modulation.
    
To evaluate the magnetic energy of the single-$q$ state, it is easier to work in the momentum space. Taking the Fourier transform of the spin wave function
\begin{align}   \label{eq:12}
\vec{S}^{(l)}_{1Q}(\mathbf{q}) &= \frac{S}{2} \left[ (\hat{e}_1^{(l)} - i \hat{e}_2^{(l)}) \delta_{\mathbf{q},\mathbf{q}'} + (\hat{e}_1^{(l)} + i \hat{e}_2^{(l)}) \delta_{\mathbf{q},-\mathbf{q}'} \right],
\end{align}
assuming the modulation vector being $\mathbf{q}'$, into $H_{\text{eff}}$ in Eq. (\ref{eq:4}), the variational energy function is obtained. These optimized parameters ($\hat{e}_1^{(1)}, \hat{e}_2^{(1)}, \hat{e}_1^{(2)}, \hat{e}_2^{(2)}$) are then used to identify the most energetically favorable spin configuration.

\begin{figure*}[!t]
\centering
\includegraphics[width=\linewidth]{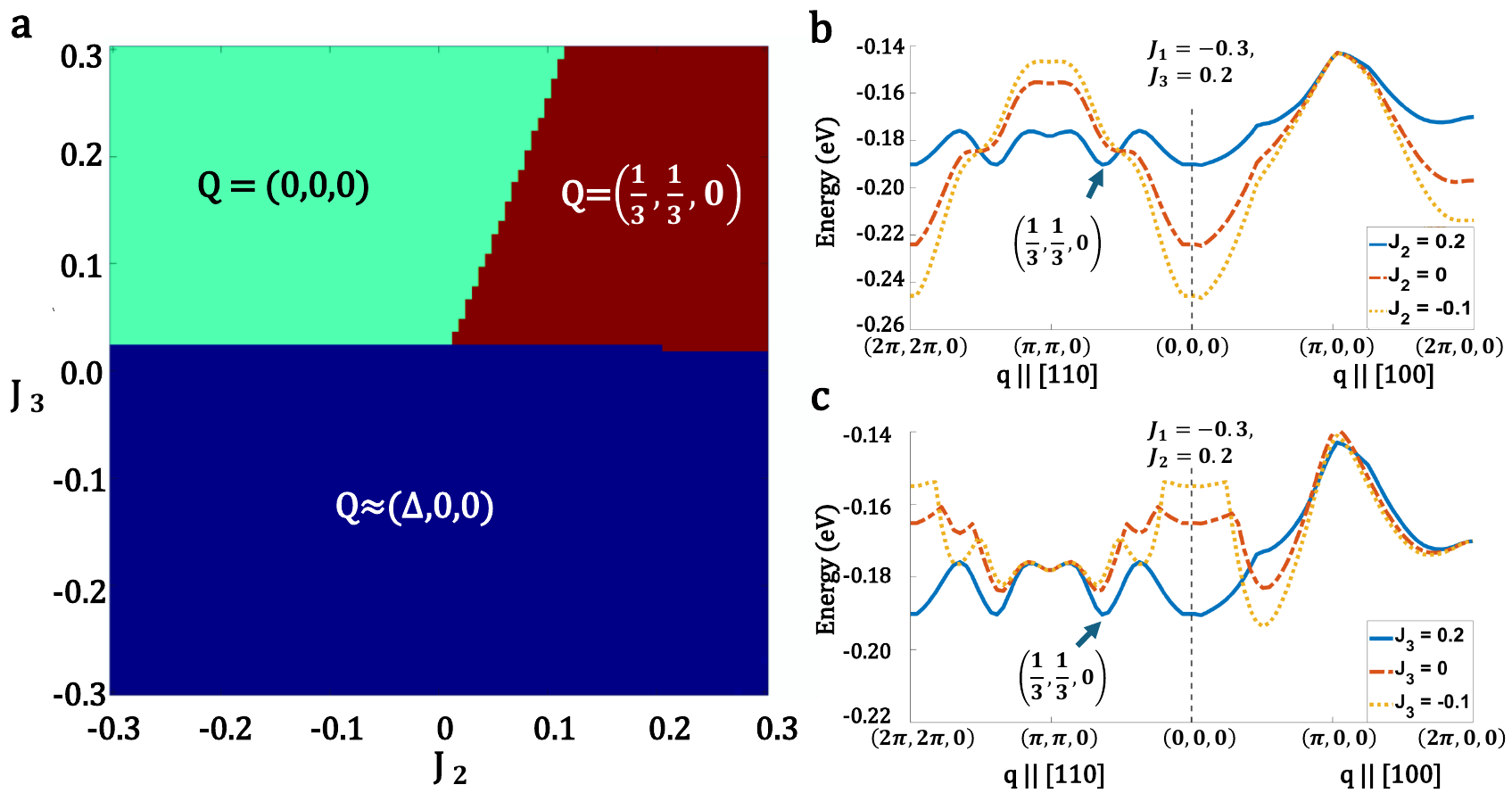}
\caption{(a) Phase diagram showing the ground-state minima as a function of $J_2$ and $J_3$, with $J_1=-0.3$ and $\Delta \approx 0.5$. Calculated dressed energy $E^{\text{RPA}}$ along [110] and [100] directions as a function of (b) $J_2$ with $J_1 =-0.3$ and $J_3 =0.2$ fixed and (c) $J_3$ with $J_1 =-0.3$ and $J_2 =0.2$ fixed. The evolution of the global minimum at $(\frac{1}{3}, \frac{1}{3}, 0)$ is clearly resolved.}
\label{fig:3}
\end{figure*}

We compare the bare and RPA susceptibilities as well as their resulting interactions. Half of the traces of the susceptibilities are shown in Fig.~\ref{fig:2}(a) and (b). (Two sublattices contribute identical values.) The RPA result reveals that interactions in $d$ electrons can enhance the effective spin interaction along the diagonal direction of $\mathbf{q}$, especially at the zone boundary near $(1/2,1/2,0)$. The diagonal interaction cannot completely determine the magnetic state, and the off-diagonal ones are also influential. The calculations show the off-diagonal terms can stabilize the $\mathbf{Q}$ order, which diverges from Fermi-surface nesting $\mathbf{q}$. (The detail of the susceptibility components are shown in Supplementary Materials.)

The competition of the diagonal and off-diagonal interactions yields the total energy. The variational energies per site of the single-$q$ states, $E^{0}$ and $E^{\text{RPA}}$, are shown in Fig.~\ref{fig:2}(c) and (d), comprising the bare and the RPA susceptibilities, respectively. The ``bare" interaction shows the minimal energy along (1,0,0), happening at $q_4$, in Fig.~\ref{fig:2}(c), while the RPA interaction shows the energy minimum exactly at $\mathbf{Q}$, marked by a circle in Fig.~\ref{fig:2}(d). The energy plot along symmetric directions is also displayed in Fig.~\ref{fig:2}(e). The PRA results in Fig.~\ref{fig:2} adopt
$J_1 = -0.3$ eV, $J_2 = 0.28$ eV, and $J_3 = 0.1$ eV. In this representive choice of parameters, the optimal magnetic order at $\mathbf{Q}$ is obtained. The choice of $J$'s indicates a ferromagnetic on-site interaction and antiferromagnetic ones for nearest-neighbor intra- and inter-sublattice spins, which might be a consequence of the superexchange mechanism.

To further explore the stability of the magnetic order at $\mathbf{Q} = \left( \frac{1}{3}, \frac{1}{3}, 0 \right)$, we map out the ground-state phase diagram as a function of $J_2$ and $J_3$ with fixed $J_1 = -0.3$, as shown in Fig.~\ref{fig:3}(a). The ordering at $\mathbf{Q}$ is stabilized only when both $J_2$ and $J_3$ are finite and positive, while other regions favor either the $\Gamma$ point or along  $(\Delta, 0, 0)$. This phase competition is more clearly resolved in Fig.~\ref{fig:3}(b) and (c), where energy dispersions along high-symmetry directions are plotted for varying $J_2$ and $J_3$, respectively. The global minimum at $\left( \frac{1}{3}, \frac{1}{3}, 0 \right)$ emerges robustly only within a narrow window of antiferromagnetic next-nearest and third-neighbor interactions. This magnetic instability is closely tied to inter-sublattice coupling, as revealed by the RPA analysis. While the intra-sublattice susceptibilities $\chi_{11}$ and $\chi_{22}$ are nearly identical, the finite inter-sublattice component $\chi_{12}$ drives the shift of the energy minimum to the vector $\mathbf{Q}$. A full analytical treatment of the susceptibility matrix and eigenvalue structure, highlighting the role of $\chi_{12}$, is provided in the Supplemental Materials.

We also examine the Luttinger–Tisza method, a widely used approximation technique for classical spin systems. Unlike the strict hard-spin constraint imposed by the mean-field method, the Luttinger–Tisza approach relaxes this condition to a softer global constraint, requiring only that the total squared spin magnitude equals the number of lattice sites. Within this framework, the eigenvalue spectrum of the effective interaction matrix formed from the combined inter- and intra-sublattice susceptibility is analyzed, and the minimum eigenvalue is identified as the ground-state energy. This method provides an efficient means of exploring the magnetic ordering tendencies of the magnetic system. In SmAlSi, the Luttinger–Tisza results closely resemble those obtained from the more rigorous mean-field ansatz, particularly when the off-diagonal elements of the interaction matrix are small. However, when inter-sublattice interactions are considered, the results begin to deviate from the mean-field predictions, reducing the method's reliability. Its accuracy tends to deteriorate in regimes where off-diagonal terms become large, leading to noticeable deviations from the true ground-state energy of the system.

\textbf{Spin structure.~}
We now discuss the spin structure of SmAlSi, derived from the optimized effective interactions corresponding to the global energy minimum. Neutron scattering experiments on SmAlSi suggest six candidate spin configurations, featuring cycloid, helical, or hybrid ordering~\cite{bib35}. In a cycloid texture, spins rotate along the direction of the modulation vector $q$, while in a helical texture, they rotate in a plane perpendicular to $q$. To identify the actual spin configuration, we analyzed both single-$q$ and double-$q$ states using susceptibility calculations and interaction-optimized parameters. The single-$q$ state involves one modulation vector, whereas the double-$q$ state includes two orthogonal modulations, enabling more intricate textures such as Skyrmions. 
The single-$q$ configuration was evaluated in momentum space and minimized the total energy density with respect to the spin orientation angles. For the double-$q$ configuration, we performed real-space simulations on an $N \times N$ lattice using a corresponding double-$q$ ansatz and applied a cutoff to suppress long-range interactions. Specifically, we compared the single modulation case with $q = (\frac{1}{3},\frac{1}{3},0)$ to the double modulation case with a linear combination of $q = (\frac{1}{3},\frac{1}{3},0)$ and $(\frac{1}{3}, -\frac{1}{3}, 0)$. Our results show that the single-$q$ configuration consistently yields a lower energy density than the double-$q$ state, indicating that the spiral single-$q$ texture is the most stable state of SmAlSi, in agreement with the hybrid structure observed in experiments.

\begin{figure}[!t]
\centering
\includegraphics[width=10 cm]{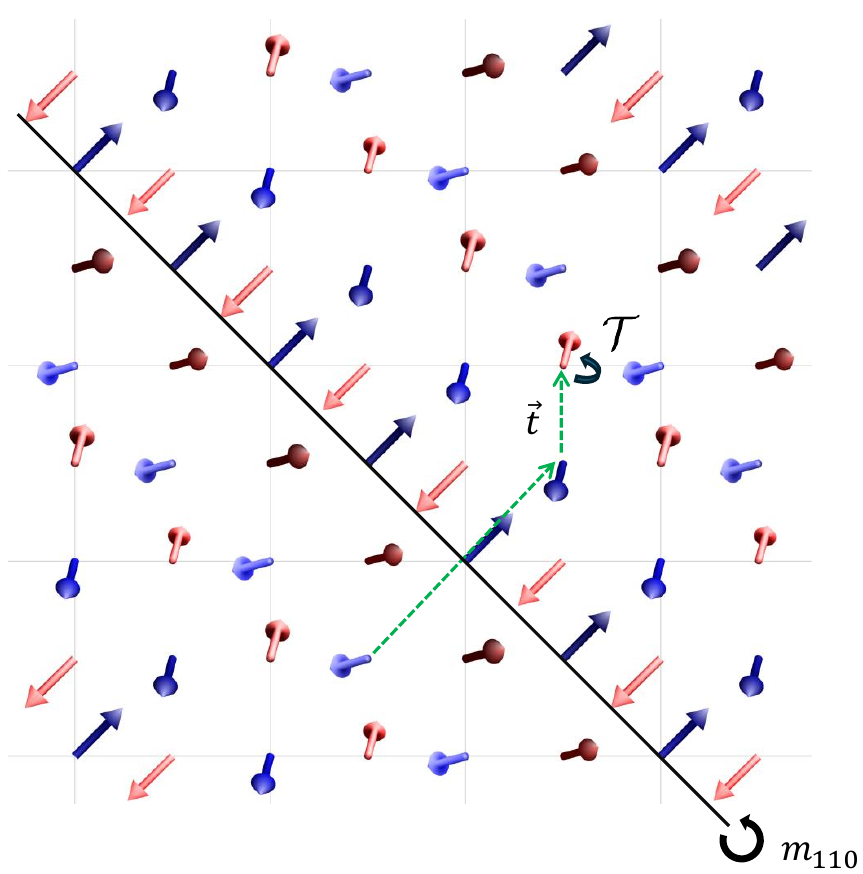}
\caption{Spin configuration of SmAlSi obtained from energy minimization. The spin textures of the two magnetic sublattices are shown in different shades of red and blue. The black line denotes the $m_{110}$ mirror plane, while the green arrows indicate a glide symmetry operation, consisting of a mirror reflection followed by a fractional lattice translation along the vector $\vec{t}$. Time-reversal symmetry $\mathcal{T}$ acts on the spin. The resulting spin configuration preserves this combined symmetry across both sublattices.
}\label{fig:4}
\end{figure}

For the single-$q$ state with the values of $J$'s as in Fig.~\ref{fig:2}, the optimized energy density is $-0.1986 $ eV ($S^2=1$ for brevity), and the corresponding magnetic order parameters on the two sublattices are
\begin{align*}
\hat{e}_1^{(1)} &= (0.7036,\ 0.7106,\ 0.0029), \\
\hat{e}_2^{(1)} &= (0.2357,\ -0.2374,\ 0.9423), \\
\hat{e}_1^{(2)} &= (-0.7106,\ -0.7036,\ 0.0029), \\
\hat{e}_2^{(2)} &= (-0.2374,\ 0.2357,\ -0.9423).
\end{align*}
The single-$q$ spin texture in real space is shown in Fig.~\ref{fig:4}. The resulting spin texture is close to a cycloidal order. 
The magnetic orders on two sublattices possess a symmetry relation, which we will elaborate on in the next section. 
The spin configuration, though complicated, still retains a symmetry. From the group theory point of view, we look for point-group symmetries in the non-magnetic crystalline group that make the modulation vector $\mathbf{Q}$ invariant or reverse its sign to $-\mathbf{Q}$. The collection of these symmetry operations forms the magnetic little cogroup. In this compound, the possible symmetry operations are $2_{001}$, $m_{1 1 0}$, $m_{1 \bar{1} 0}$. If none of them exists, the magnetic order has no symmetry except translation. 

The observed symmetry in our solution, Fig.~\ref{fig:4}, is $\mathcal{G}' = \mathcal{T} \mathcal{G}_1$, where $\mathcal{T}$ denotes time-reversal symmetry and $\mathcal{G}_1$ is the glide plane. This belongs to type-III magnetic space group, denoted $Cc'$ (no. 9.39). We exemplify in Fig.~\ref{fig:4} the symmetry with a chosen blue spin from the left bottom, through the mirror reflection $m_{[110]}$, then a fractional translation $\vec{t}=\left(0,\frac{1}{2},\frac{1}{4} \right)$, and finally time-reversal $\mathcal{T}$ to the top red spin. 

It's known that helical or cycloidal spirals within a single-sublattice lattice are chiral, thus breaking mirror symmetries. However, a cycloidal spiral can still exhibit a combined $\mathcal{T} M $ symmetry, where $M$ is a mirror plane perpendicular to its modulation vector. Surprisingly, this characteristic is partially manifested in our two-sublattice system, despite the replacement of the mirror plane by the glide plane that exchanges two sublattices. Referring to the spin texture presented in Fig.~\ref{fig:4},  the spins within each sublattice are approximately confined to the $[1 \bar{1} 0]$ plane. Therefore, we conclude that the magnetic state closely approximates a cycloidal spiral.

\textbf{Summary and disucssion.~}
We have investigated the magnetic ground state of SmAlSi by considering short-range superexchange interactions between $f$ states, mediated by itinerant $d$ electrons near Weyl nodes. Although the $f$ bands lie well below the Fermi level, the presence of Weyl nodes and partial Fermi surface nesting satisfies the conditions for Weyl-mediated RKKY interactions. Using a first-principles Wannier tight-binding model, we incorporated both $f$-$d$ and $d$-$d$ couplings and calculated both bare and dressed susceptibilities that account for Weyl-mediated RKKY interactions. Our effective spin model includes onsite, nearest-neighbor, and inter-sublattice interactions ($J_1$, $J_2$, $J_3$). For appropriate interaction strengths, the magnetic energy minimum occurs at wavevector $\mathbf{Q}=(\frac{1}{3}, \frac{1}{3}, 0)$, consistent with neutron scattering experiments. Both mean-field and Luttinger–Tisza analyses support a single-$q$ spiral ground state. The resulting spin texture is a close to cycloidal structure, constrained by the magnetic symmetry of the system.  

Although Fermi surface nesting is observed, it does not coincide with the experimentally observed $\mathbf{Q}$ vector, suggesting that Weyl-mediated interactions rather than conventional nesting are responsible for the magnetic ordering in SmAlSi. These results provide a microscopic framework for understanding magnetism in SmAlSi and underscore the broader role of Weyl fermions in stabilizing unconventional spin textures. While Weyl nodes are well known for driving anomalous transport and optical phenomena, our findings demonstrate that they are also important in mediating magnetic interactions that lead to novel spin structures, opening new avenues for fundamental research and potential applications in spintronics and quantum technologies.
	
\section*{Acknowledgments}
The work at NSYSU is supported by the NSTC-AFOSR Taiwan program on Topological and Nanostructured Materials, Grant No. 110-2124-M-110-002-MY3 and NSTC112-2112-M-110-012. SMH also thanks the support from NCTS in Taiwan. The work at TIFR Mumbai is supported by the Department of Atomic Energy, Government of India, under Project No. 12-R\&D-TFR-5.10-0100, and benefited from the computational resources of TIFR Mumbai.

\end{document}